# POLYSTYRENE-BASED SCINTILLATION DETECTOR STUDIES IN ASTROPARTICLE PHYSICS AND MEDICAL PHYSICS.

## Dr. Sonali Bhatnagar,

*Department of Physics and Computer Science, Faculty of Science, Dayalbagh Educational Institute, Agra, India*

Email: sonalibhatnagar@dei.ac.in

## ABSTRACT

DEASA (Dayalbagh Educational Air Shower Array) consists of eight plastic scintillators each with an area of 1 square meter. The cosmic ray showers have been simulated in CORSIKA [1] for the different primary particles in the energy range of $10^{14}$- $10^{15}$ eV. The longitudinal and lateral profiles have been studied for Agra.

The real-life applications of cosmic ray particles in space have been studied to protect the astronaut from the galactic cosmic rays [2]. A plastic scintillation detector is simulated in Geant4 to study applications in hadron and carbon ion therapy [3]. The proton and carbon beam are simulated through the tumour region to study the stopping power and depth dose distribution for different organs. The energy range for each study is optimized and the Bragg curve is then interpreted with Bragg peak position and range.

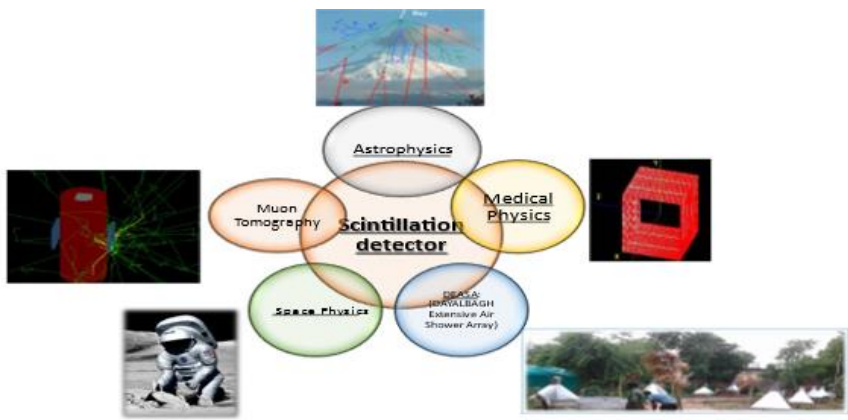

# 1. INTRODUCTION

A scintillation detector is made from materials that emit small flashes of light or scintillation when they are struck by ionizing radiation. Scintillation is a property of luminescence which is triggered by ionizing radiation whereas luminescence is spontaneous conversion of light which is not due to heat. The first scintillator was invented by Sir William Crookes in 1903 it was a ZnS plate which emitted scintillation light when struck by radiation. He observed the scintillation light under a microscope when radium bromide salt was spilt on a ZnS screen. This light could be seen with the naked eye in a darkened room and so the first ZnS scintillator called Spinthariscope was born. Later this detection technique was responsible for the discovery of the nucleus in the experiments by Rutherford and Geiger-Marsden in 1909 during the experiments of alpha particle scattering on Au foil. Coming to the discovery of present-day scintillator detectors, the naked eye measurement was replaced by a photomultiplier tube by Curran and Baker in 1944. The emitted light is detected, amplified and transformed into an electrical signal that can be measured. This detected signal gives information on the properties of the ionizing radiation.

These detectors find use in nuclear physics experiments such as nuclear reactions and others to study nuclear structure and decays [4]. Based on these studies the requirements for scintillating materials are to be sensitive to the radiation detected, should have a linear energy response with a reasonable efficiency, fast response and the attribute to discriminate different types of radiation. Another important factor in this is the stopping power which describes the capability to stop or completely absorb the incident radiation. The stopping power for charged particles is due to Coulomb interaction which increases with the atomic number of the material and its thickness. For neutral particles, the stopping power is based on the photoelectric effect or Compton interaction or pair production. Mostly in nuclear physics two types of scintillators are used- inorganic scintillators in the form of crystals and organic scintillators in the form of crystals, liquids, and plastics. Nowadays noble gas scintillators are also being used in Particle and astroparticle physics.

In the inorganic scintillator, the mechanism comprises absorption where the energy of the ionizing radiation is used to create primary electrons in the valence band and inner shell holes. After this, the energetic electrons get multiplied by radiative decay (secondary X-rays), non-radiative decay (Auger electron) and inelastic scattering of electrons and holes. This results in the thermalization of electrons (bottom of the conduction band) and holes (top of the valence band) and migration of the energy carriers to the luminescent site (away from the site of absorption). Then the relaxation happens which is the emission of energy or scintillation by the excited emission center and the luminescent site is an impurity atom or a particular crystal structure. The corresponding wavelength is in the ultra-violet or visible range and doping in crystals is necessary for shifting the wavelength to the visible region. In the case of organic scintillator, the delocalized $\pi$ electrons give rise to excitations and the de-excitation process of electrons generates the luminescence (light). The fluorescence from the singlet states is faster of order nanosec with wavelengths in deep ultraviolet regions and the phosphorescence from the triplet states is comparatively slower (microsec). The absorption and emission wavelengths need to be different to reduce self-absorption and increase transparency in the scintillator. This is achieved through doping with more than one material in the form of a primary shifter and secondary shifter as described in Figure 1.

|  | solvent | secondary fluor | tertiary fluor |
|---|---|---|---|
| Liquid scintillators | Benzene<br>Toluene<br>Xylene | p-terphenyl<br>DPO<br>PBD | POPOP<br>BBO<br>BPO |
| Plastic scintillators | Polyvinylbenzene<br>Polyvinyltoluene<br>Polystyrene | p-terphenyl<br>DPO<br>PBD | POPOP<br>TBP<br>BBO<br>DPS |

Figure 1 Different scintillators.

The organic scintillators cannot be used for high energy radiation and heavy charged particles and are mostly useful for pulsed shape discrimination. Finally, for comparison studies, the inorganic scintillators (IOS) have a high density ($>8gm/cm^2$) implying higher stopping power as compared to 1 $gm/cm^2$ for organic scintillators (OS) resulting in lower stopping power. The IOS have better energy resolution of order 60,000 photons per MeV as compared to 40,000 photons per MeV. The IOS cannot be fast detectors (except $BaF_2$) whereas OS have decay time of order of nanosec. The IOS have mostly linear energy output giving the best energy resolutions whereas the organic ones are best for fast signals.

The choice of detector in an experimental setup is based on optimizing the efficiency and energy resolution for energy spectroscopy optimizing the efficiency and time resolution for timing spectroscopy or optimizing the stopping power for calorimetric measurements. The IOS is used for particle lifetime measurements, and for high-resolution gamma detectors and liquid OS are used for the detection of neutrons. Thus, the scintillators are among the important detectors used for many cosmic rays [5] and high energy experiments such as balloon experiments to detect cosmic rays in TRACER, TIGER and CREAM [6]. The liquid scintillators have been used in the neutrino experiments of the next generation including Ice Cube in Antarctica. Plastic scintillators due to their design flexibility are being used in almost all ground-based arrays such as KASCADE [7], Telescope array [8], Tibet array [9] and GRAPES3[10].

DEASA array at the Dayalbagh Educational Institute (27 °22'N, 78° E, 170 m a.s.l, 1013 g/cm$^2$) has eight plastic scintillation detectors installed in two square lattices with 8m separation. The detectors with an area of 100 x 100 cm$^2$ are connected to basic electronic NIM (nuclear instrumentation modules) which can discriminate the PMT output from the noise threshold. DEASA aims to study air showers having energies around the knee region of cosmic rays. This paper outlines the applications studied based on plastic scintillation detectors in real life. These applications are in astrophysics where the muon telescope was used to measure muon flux and interpret it with temperature, pressure and angular dependency at Agra. In the second section, the Monte Carlo simulations of detectors are done for oncology studies.

## 2. APPLICATIONS

DEASA (Dayalbagh Educational Air Shower Array) is an experiment as shown in Figure 2 where the users are undergraduate and post-graduate students in physics. Since this is the only air shower array in northern India, hence exposure to astrophysical studies is important. The educational activities for the UG and PG students involve cosmic ray activities, outreach in the form of workshops and citizen science projects.

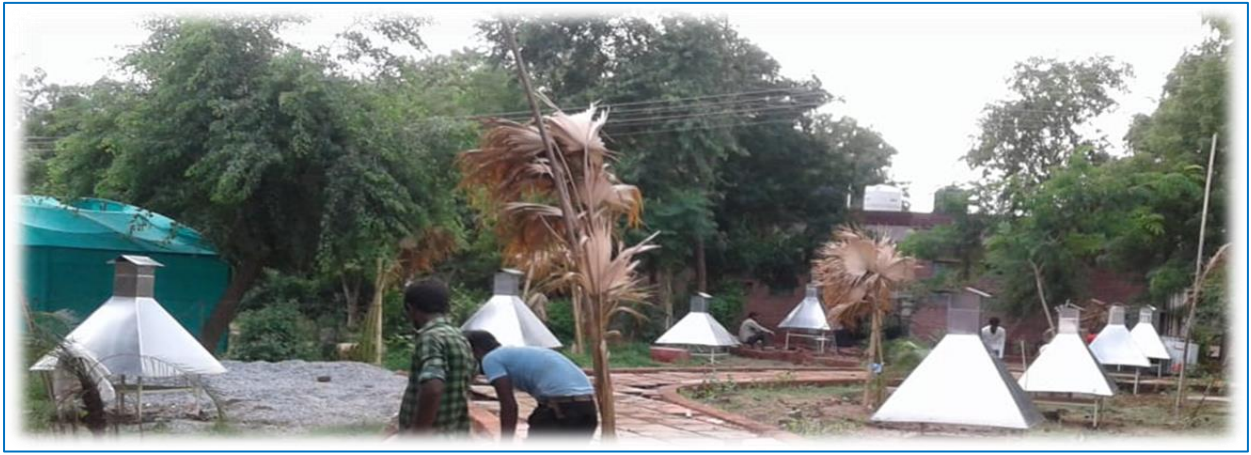

*Figure 2 DAYALBAGH EDUCATIONAL AIR SHOWER ARRAY*

## 2.1 ASTROPHYSICAL STUDIES OF DEASA

The prototype detectors of the mini array are a single-fibre and double-fiber polystyrene plastic scintillator with single wavelength shifting fibres and double WLS fibres respectively in small grooves on them. These are also called paddles being 23.5 x 24 $cm^2$ in area and kept in five different positions to perform different studies. The plastic detectors were made, cut, polished and then wrapped in Tyvek sheet, put in the aluminium box and shipped to Agra. These were calibrated and checked for light leakage, and operating voltage was found based on a coincidence window of 100 nsec. The NIM comprised of discriminator, logic unit and a 4-channel counter for recording the muon flux in predefined time [11]. The telescope solid angle was calculated as 0.15 radians with no separation and decreased further as a function of separation. The muon flux was observed inside, and outside the laboratory and then on the roof of the laboratory as shown in Figure 3.

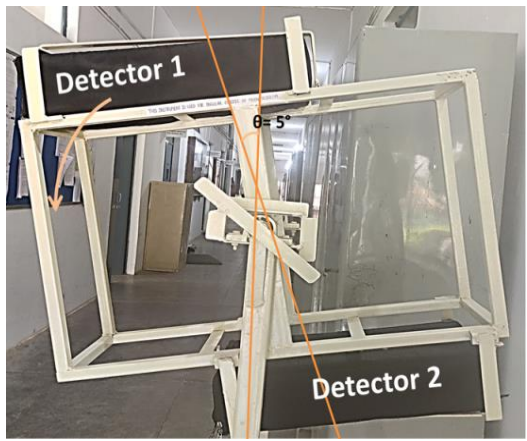

*Figure 3 Muon telescope*

The studies are plotted and compared with the change in flux of muons measured by varying distance between the paddles in all three locations. The results of this paper showed variation of secondary air shower particles in the data on ground and on the roof. Another study published by the laboratory is studying atmospheric effects on muon flux and decoherence curve is plotted. In this work simulation of the paddles was done in geant4 and effects of multiple scattering on the interaction lengths of secondary particles of different energies are plotted [12]. These studies have been simulating high energy secondaries observed on the ground in geant4 using a UNIFIED model and different models for muon interactions depending on their energies. The response of plastic scintillators to muons of energy 110 MeV -180 MeV is shown. Recently the laboratory has published a preliminary CORSIKA simulation [1] with the longitudinal and lateral profile of air showers.

## 2.2 MEDICAL PHYSICS

Hadrontherapy is an advanced technique of external radiation therapy, using either low-LET ions such as protons or high-LET particles such as carbon ions. In contrast with photons, ions have the advantage of stopping at a given depth and delivering a maximum dose in the distal part, called Bragg-peak, where the tumour is located. Moreover, high-LET ions have enhanced radiobiological properties, which allow for treating radio-resistant tumors. It is developed to destroy cancerous tumors by the use of ionizing particles along with maintaining the functioning of adjacent healthy tissues. The emission of energy from radiation and its propagation through space or a material medium is different from particle radiation. The term particle radiation is meant for the energy transmitted in cells with considerable mass and momentum. The relationship among these particles can only be explained through the Standard Model of fundamental particles which may not be a complete theory, but it helps in framework of theoretical and experimental advances.

While compared to X-rays, the charged particle beams have a finite range of ballistics which helps in defining finest depth within the tissue. Both proton and $^{12}$C beams deposit their maximum dose located at the end of the path. This results in a Bragg Peak where its location depends on the beam energy, tissue density, and its composition. The proton beam gradually slows down by energy loss undergoing Multiple Coulomb Scattering with the material. The consistency of dose deposition may also be influenced by inhomogeneous materials present in the path of beam direction. These inhomogenities results in degradation of Bragg peak and its range. This uncertainty in dose distribution can be determined from Full Width Half Maximum (distance between the entrance surface of beam and distal FWHM) which has a large impact on treatment system because maximum dose of proton may be delivered to normal tissue. The point at 80% of energy deposition results in particle range during beam irradiation. Another parameter, penumbra width gives the resultant absorbed dose in distal region of Bragg peak. At low energy range (<100 MeV), proton beams is a key issue in various experimental domain such as oncological radiotherapy, radiation protection, space science and optimization of radiation monitoring detectors. A number of medical physicists are making collaboration with Stanford Linear Accelerator Center (SLAC) for working on Geant4 applications, especially for the treatment of tumors in hadron therapy and brachytherapy.

GEANT4 (GEometry ANd Tracking 4) is a versatile C++ Monte Carlo simulation toolkit developed at CERN. It has been initially created for high energy applications up to about 100 TeV and now allows for very low energy applications down to a few eV. As a particle physics simulation toolkit, GEANT4 simulates all kind of particles, including exotic particles. A large number of processes, models, cross-sections and simulation parameters are available and they have to be carefully selected depending on the application. Available models are growing continuously and are historically divided into three categories: data-driven, parameterized and theory-based models. The data-driven approach is considered to be the optimal way of modelling, but models are not strictly data-driven, theory-based or parameterized: for instance, a data-driven model can be parameterized for some parts if data is missing. For medical applications, various validation studies have been proposed, covering the validation of photon and electron physics for conventional radiotherapy and the validation of electromagnetic and nuclear interactions for carbon ion and proton therapy applications. The main task of a medical physicist is to make sure the right dose is delivered in the right place. Treatment planning systems (TPS) used to plan radiotherapy treatments are fast, but with limited dose accuracy in some particular cases. On the other hand, Monte Carlo simulations are slower but are considered to be the reference for dose calculation accuracy. Therefore, Monte Carlo is a valuable tool to benchmark TPS.

The stopping power of proton beams in various materials like brass, aluminium, copper, water, lead, plexi glass and scintillation detectors is computed within an energy range of 0.1 to $10^5$ MeV. The physical distribution of dose curves of proton beam is studied from Geant4 simulation of the setup installed at INFN Catania for curing ocular tumours. The significance of the energy spectrum of secondaries generated during irradiation of 60 MeV proton beam is also observed from the study of Radioactive Biological Efficiency (RBE) and Spread Out in Bragg Peak (SOBP)[2]. The dose deposition of a passive proton beam of 60 – 240 MeV in a human phantom is determined. The Bragg peak and its parameters are calculated and verified with the results of other groups. The effect of the various mediums in tumours has been done by numerous reports that help in finding the ability of beam density to determine the dose equivalent required. Adipose, bone and soft tissue materials are simulated as inhomogeneous mediums along their positioning of the tumour region are also designed. The energy range which is necessary in curing a tumor corresponding to its volume is optimized. The parameters which have their impact on the Bragg curve are also computed. A substantial error in the Bragg peak position is observed which is approximately -0.36% for the muscular skeleton, and -0.44% for soft tissue. The major effect on the dose equivalent of proton beam which is necessary for the treatment of tumors in human phantom at different ages is studied. Distinctive volumes of human phantoms at various ages were also simulated. The relation between dose equivalent and absorbed dose as a function of depth to target volume is computed. A dependency of the dose deposition on the patient's age is also observed i.e. at a younger age, a higher amount of dose is necessary, while it gradually decreases in the other cases i.e. 10 years, 15 years and adult.

## 3. CONCLUSION

The applications of polymer-based plastic scintillation detectors is vast. These detectors are an integral part of the Astroparticle physics experiments like balloon experiments, and space flights. These detectors are aiding medical physicists in Hadrontherapy, gamma cameras and simulations of carbon ion beams. These detectors are part of the modulators to reach the tumour. These scintillators are important for space simulations and material studies for absorption of galactic cosmic rays. Finally, these scintillation detectors play a very important role in the nuclear imaging technique for identification of nuclear waste and how radioactive the material is. These materials are also being exploited in fundamental particle research such as neutrino experiments and dark matter experimental search.

**ACKNOWLEDGEMENT**

The author acknowledges the Dayalbagh Educational Institute for funding the Dayalbagh Educational Air Shower Array in the Faculty of Science.

## Short Biography of Dr Sonali Bhatnagar

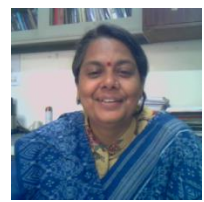

Dr. Sonali Bhatnagar is a Senior Assistant Professor of physics at Dayalbagh Educational Institute, Agra. She received her Ph.D. degree from Punjab University, Chandigarh. Her current area of expertise is in high energy physics, has set up a mini array for studying secondary cosmic rays at Agra called Dayalbagh Educational Air Shower Array (DEASA) in collaboration with T.I.F.R., Mumbai. She was awarded the

Young Scientist Award FOR WOMEN IN SCIENCE at the FMT 2020 Conference, by the Department of Physics, School of Applied Sciences, Kalinga Institute of Industrial Technology (KIIT). She has published around 45 peer-reviewed articles.